\documentclass[aps,prb,twocolumn,groupedaddress,showpacs]{revtex4}
\usepackage{graphicx}

\begin{document}

\preprint{}

\title{ Electronic Structure of Three-Dimensional Superlattices\\
Subject to Tilted Magnetic Fields}
\author{N.\ A.\ Goncharuk, L.\ Smr\v{c}ka, J.\ Ku\v{c}era}
\author{K.\ V\'yborn\'y} 

\altaffiliation[present address:]{ Institute of Theoretical Physics, 
University of Hamburg, Jungiusstr. 9, 20355 Hamburg, Germany.}

\affiliation{Institute of Physics, Academy of Science of the Czech
Republic, Cukrovarnick\'{a} 10, 162 53 Praha  6, Czech Republic}


%\author{}
%\thanks{}
\affiliation{}


\date{\today}

%=====================================================================
\begin{abstract}

Full quantum\discretionary{-}{-}{-}mechanical description of 
electrons moving in 3D
structures with unidirectional periodic modulation subject to tilted
magnetic fields requires an extensive numerical calculation. To
understand magneto-oscillations in such systems it is in many cases
sufficient to use the quasi\discretionary{-}{-}{-}classical approach, 
in which the zero-magnetic-field Fermi surface is considered as a
magnetic-field-independent rigid body in $\vec{k}$-space and periods
of oscillations are related to extremal
cross-sections of the Fermi surface cut by planes perpendicular to the
magnetic-field direction.  We point out cases where the
quasi\discretionary{-}{-}{-}classical treatment fails and 
propose a simple tight-binding 
fully-quantum\discretionary{-}{-}{-}mechanical model of the 
superlattice electronic structure.

\end{abstract}
%=====================================================================

\pacs{73.21.Cd, 73.20.At, 03.65.Sq, 03.65.-w}
%\keywords{}

\maketitle

%\section{}

%%%%%%%%%%%%%%%%%%%%%%%%%%%%%%%%%%%%%%%%%%%%%%%%%%%%%%%%%%%%%%%%%%%%%%%
\section{Introduction\label{Intro}}
%%%%%%%%%%%%%%%%%%%%%%%%%%%%%%%%%%%%
Esaki and Tsu predicted Bloch oscillations in semiconductor
superlattices in 1970 \cite{Esaki} and since then extensive studies of
electron dynamics in these structures have been carried out. The
review of research till 1987 was given by Mann \cite{maan}, who also
discussed the quantization of band-structure by a magnetic field
parallel to the layers.

Presently, various aspects of the superlattice electronic properties
are investigated.

In quasi\discretionary{-}{-}{-}two\discretionary{-}{-}{-}dimensional 
(2D) layered organic conductors two distinct
fundamental concepts of electron interlayer transport are considered,
the coherent and the incoherent.  The comparison of both approaches was
presented by McKenzie and Moses \cite{Moses1,Moses2}. It has been
demonstrated that the dependence of the interlayer magnetoresistance
in aforementioned natural structures on the direction of the magnetic
field is identical for both models except the case of a field almost
parallel to the layers, when Yamaji oscillations \cite{Yamaji} can
occur.  An explanation of magnetoresistance angular effects (Yamaji
oscillations) observed in layered organic conductors has been given in
the framework of the incoherent model of interlayer coupling in
anisotropic multilayer systems \cite{Osada1, Osada2, Osada3}.
 
In the semiconductor superlattices, the band profile of which is
formed by a periodic sequence of quantum wells, general belief is
that the electron motion along the growth direction is coherent
and governed by the Bloch theorem.  As result, the electrons can
move freely parallel to the plane of wells and their motion in the
growth direction is described by minibands.

Here we consider short-period superlattices with only the lowest
electron miniband occupied.  In such a case, the superlattice
electronic structure is close to the 
three\discretionary{-}{-}{-}dimensional (3D) electron
system when the Fermi energy, $E_{F}$, lies below the top of the
miniband and a Fermi surface forms a closed oval in the first
Brillouin zone. When the Fermi energy coincides with the top of the
miniband, the Fermi surface consists of the chain of stretched ovals
``kissing'' on Brillouin zone borders in the repeated zone scheme.
For Fermi energies lying above the top, the Fermi surface is open and
acquires the form of a corrugated cylinder. In the limiting case of
impenetrable barriers the miniband width is reduced to zero and the
superlattice is converted into a multiple 2D electron layers.
We also limit ourselves to investigation of the electron structure
magneto-oscillations and will not describe a specific property as 
magnetization (de Haas-van Alphen oscillations) or magnetoresistance
(Shubnikov-de Haas oscillations).

To distinguish between 2D and 3D electron systems in superlattices,
tilted magnetic fields, $\vec{B}$, are used.  In 3D systems the
magneto-oscillations are observed for an arbitrary
magnetic field orientation, whereas in 2D systems the oscillations are
determined only by a perpendicular field component, $B_{\perp}$, and
disappear in the in-plane fields, $B_y$. The tilted magnetic
field configuration was used e.g.  to confirm the 3D nature of a
semiconductor superlattice on which the existence of quantum Hall
effect in 3D structures was proven \nolinebreak[3] \cite{Stormer}.

The quasi\discretionary{-}{-}{-}classical approach to 
interpretation of magnetotransport
experiments relies on the Onsager-Lifshitz quantization rule
\cite{on,Ashcroft}.  The theory states that magneto-oscillations are
periodic in $1/B$, the period of oscillations is determined by the
extremal cross-sections of the Fermi surface perpendicular to the
direction of the applied magnetic fields. A number of extremal
cross-sections can be examined and the shape of 3D Fermi surface
reconstructed by tilting the sample in the magnetic field. The
quasi\discretionary{-}{-}{-}classical approach is also 
employed in studies of chaos associated with instability 
of electron orbits in the presence of a
tilted magnetic field \cite{Eaves1, Eaves2, Osada4, Eaves3}.

The theory of magnetic breakdown \cite{Falicov,Blount, Chambers} goes
beyond the quasi\discretionary{-}{-}{-}classical 
approximation by taking into account
tunneling between eigenstates evaluated 
quasi\discretionary{-}{-}{-}classically (by the WKB
method), i.e. it is implicitly assumed that the states with high
quantum numbers are involved.

The experimental evidence of deviations from the 
quasi\discretionary{-}{-}{-}classical
interpretation of data measured in tilted magnetic fields on
semiconductor superlattices has been reported in \cite{Jaschinski,
Nachtwei} and \cite{ Kawamura_1,Kawamura_2}. The reason is attributed to the
in-plane component which is supposed to reduce the 
tunneling of electrons between wells when their separation is
comparable with the in-plane magnetic field length,
$l_y=\sqrt{\hbar / |e|B_y }$, as first proposed by
Dingle \cite{Dingle} in 1978.

We will study this problem theoretically using a simple tight-binding,
fully-quantum\discretionary{-}{-}{-}mechanical model of the 
superlattice electronic structure in which the generally 
three\discretionary{-}{-}{-}dimensional Schr\"{o}dinger equation reduces to an 
one\discretionary{-}{-}{-}dimensional differential equation.

Our approach is an extension of the model developed in 1992 by Hu and
MacDonald \cite{Hu} for electron bilayers subject to tilted magnetic
fields, which was since then many times successfully applied to
semi-quantitative interpretation of the experimental data. Two basic
approximations are employed: (i) The electron layers confined in
quantum wells are strictly two\discretionary{-}{-}{-}dimensional, 
i.e. there is no influence
of the magnetic-field-in-plane component on the individual layer. (ii)
The barrier width and the barrier height are represented by a single
coupling parameter $t$.  Thus the problem is characterized by two
parameters, the hopping integral $t$ and the interlayer distance
$d_z$.

In Sec. \ref{Tight} we briefly summarize the textbook results
(obtained with the aid of the above model) for the electronic
structure of short-period superlattices in zero magnetic field. The
discussion of the electronic structure in tilted magnetic fields is
opened in Sec. \ref{Quasi} by presentation of the 
quasi\discretionary{-}{-}{-}classical
results in the form appropriate for comparison with the subsequent
quantum\discretionary{-}{-}{-}mechanical treatment in \ref{Quant}, 
which represents the central part of this paper. Sec. \ref{In-plane} 
is devoted to the case of strictly in-plane magnetic fields not 
covered by the previous discussion.  Numerical results are 
offered in Sec. \ref{Num}, followed by concluding remarks in Sec. \ref{Concl}.

%%%%%%%%%%%%%%%%%%%%%%%%%%%%%%%%%%%%%%%%%%%%%%%%%%%%%%%%%%%%%%%%%%%%%%%%%
\section{A tight-binding  miniband \label{Tight}}
%%%%%%%%%%%%%%%%%%%%%%%%%%%%%%%%%%%%%%%%%%%%%%%%
A tight-binding model of minibands in 3D superlattices can be found
e.g. in \cite{bastard}. In this model, a superlattice is formed by a
periodic sequence of quantum wells separated by barriers, with the
potential energy $V(z)$ written as a sum of potential energies
$V_b(z)$ of individual wells,
\begin{equation}
V(z)  =  \sum_{j} V_b(z-Z_j).
\end{equation}
Here $Z_j = jd_z$, $j$ is an integer and $d_z$ is a period of the
superlattice. Then the $z$-dependent part of the Hamiltonian, $H_z$,
reads
\begin{equation}
H_z = \frac{p_z^2}{2m} + V(z).
\label{hamz}
\end{equation}
For the narrow wells we considered only the lowest electron miniband
of the superlattice. Only the ground states $|\chi_b(z-Z_j)\rangle$ of
individual wells enter our model. Their eigenenergies are taken as an
origin of the energy scale.  The eigenenergies of excited states are
assumed to lie well above them and their presence is neglected.

In such structures the plane wave $\exp(ik_xx+ik_yy)$ describes
electrons moving in the $x y$-plane, the electron motion in the 
$z$-direction is mediated by tunneling through the barriers between wells.
A wavefunction $\chi_{k_z}(z)$ describing the miniband is the Bloch
sum
\begin{equation}
\chi_{k_z}(z) = \sum_je^{ik_z Z_j} \chi_b(z-Z_j).
\label{bloch}
\end{equation}
The diagonal matrix elements of $H_z$ are equal to zero in the basis
of ground states $|\chi_{b,j}\rangle\equiv|\chi_b(z-Z_j)\rangle$,
$\langle\chi_{b,j}|H_z|\chi_{b,j}\rangle=0$. Only the hopping
integrals $\langle\chi_{b,i}|V(z)|\chi_{b,j}\rangle=-t
\,\delta_{j,i\pm 1}$ are nonzero if we assume the nearest-neighbor
interaction between the individual wells.  As the hopping integrals
are negative, $t$ is a positive constant, i.e.  in our notation
$t=|t|$.
%
%^^^^^^^^^^^^^^^^^^
\begin{figure}[t]
\includegraphics[width=\linewidth]{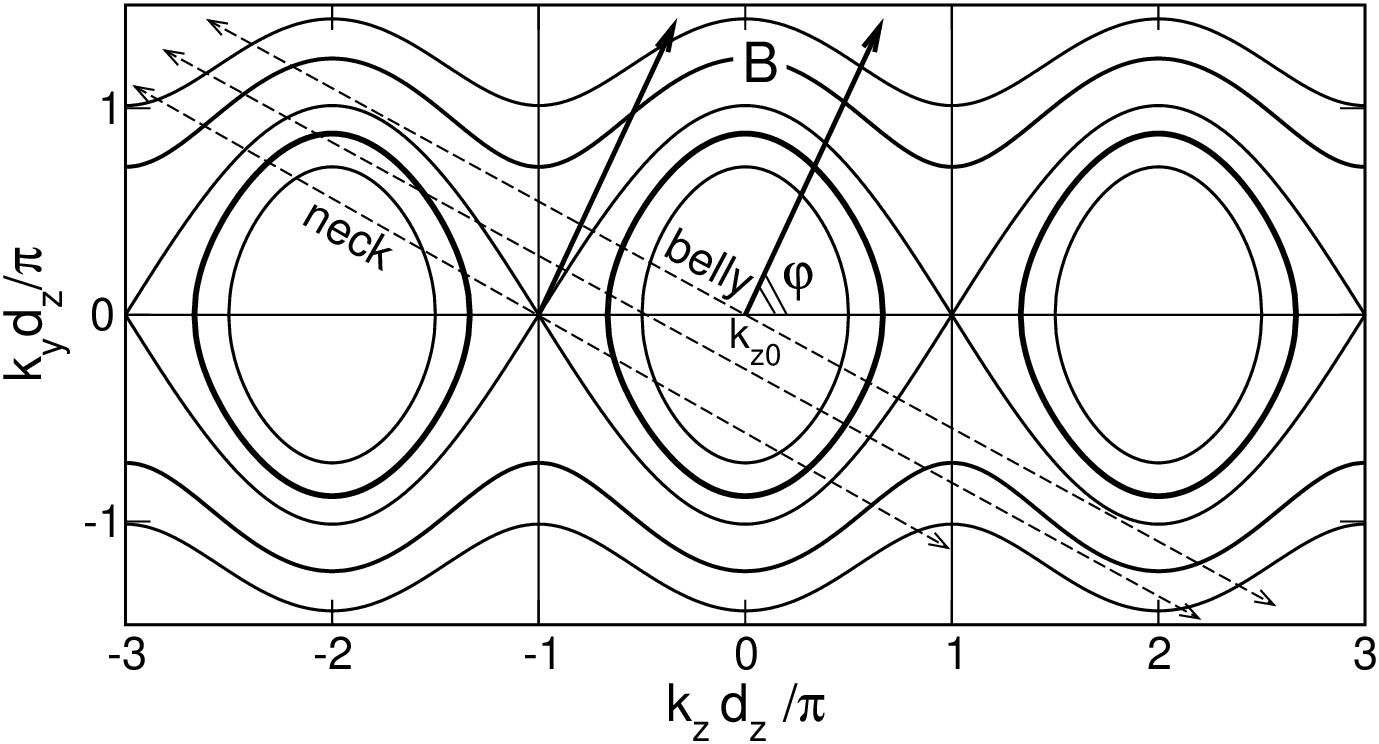}
\caption{\label{Fig1}Fermi surfaces of a superlattice plotted for
Fermi energies lying 5 meV, 7.5 meV, 10 meV, 15 meV, and 20 meV above
the miniband bottom.  The tilt angle $\varphi$ is 65$^{\circ}$. }
\end{figure}
%^^^^^^^^^^^^^^^^^^

The matrix equation which determines the $z$-dependent part of the
eigenenergies reads
\begin{equation}
\langle\chi_b(z-Z_j)|E-H_z|\chi_{k_z}(z)\rangle =0.
\end{equation}
The resulting dispersion relation $E(k_z)$ of a miniband has a simple
cosine form,
\begin{equation}
E(k_z)  = -2t\cos(k_zd_z),
\end{equation}
which depends on two parameters, $t$ and $d_z$.  Note that the
eigenfunctions $\chi_{k_z}(z)$, described by equation (\ref{bloch}),
are fully determined by the superlattice translation symmetry.

The energy spectrum of the 3D electron motion is 
composed of $E(k_z)$ and the energy of 
the free motion in the $x y$-plane:
\begin{equation}
E(\vec{k}) = \frac{\hbar^2}{2m}(k_x^2+k_y^2) - 2t \cos(k_zd_z).
\label{disp}
\end{equation}
The period of the superlattice $d_z$ determines the size of the
Brillouin zone, defined by $-\pi/d_z < k_z <\pi/d_z$. For the
Fermi energy in the range of miniband energies, $-2t < E_F <
2t$, the Fermi surface has a closed semi-elliptic shape, for $E_F >
2t$ it is an open corrugated cylinder. Examples of
constant energy surfaces are shown in Fig.~\ref{Fig1} for a superlattice with
the period $d_z = 24$ nm, the miniband width $4t = 10$ meV and the
electron effective mass $m =$ 0.067. (These parameter values will
be maintained through the whole paper unless stated otherwise.)  With
vanishing $t$, the system is transformed into a sequence of
independent 2D electron layers and the Fermi surface becomes a smooth
cylinder for any $E_F$.

%%%%%%%%%%%%%%%%%%%%%%%%%%%%%%%%%%%%%%%%%%%%%%%%%%%%%%%%%%%%%%%%%%%%%%%%%
\section{Tilted magnetic fields \label{Tilted}}
%%%%%%%%%%%%%%%%%%%%%%%%%%%%%%%%
\subsection{Quasi-classical approach\label{Quasi}}
%%%%%%%%%%%%%%%%%%%%%%%%%%%%%%%%%%%%%
The standard quasi\discretionary{-}{-}{-}classical approach 
to the electronic structure of
superlattices in tilted magnetic fields and to the interpretation of
related experiments is based on the  Onsager-Lifshitz
quantization rule,
\begin{equation}
A_k = \frac{2\pi|e|B}{\hbar}(n+\textstyle\frac{1}{2}),
\label{ons}
\end{equation}
where $A_k$ is an area of the extremal cross-section of the Fermi
surface, perpendicular to the direction of the applied magnetic field
$\vec{B}$. The plane perpendicular to $\vec{B}$, which cuts the 
$k_z$-axes at $k_z=k_{z0}$, is described by 
\begin{equation}
k_z = \frac{B_y}{B_z}\,k_y - k_{z0}.
\label{kz}
\end{equation}
There are two extremal cross-sections, $k_{z0}=\pi/d_z$ corresponds to
the ``neck'' orbit and $k_{z0}=0$ to the ``belly'' orbit.  In the real
space, the electrons move along the orbits which have the same shapes
as the contours of the cross-sections, but are rotated by 90$^{\circ}$
and scaled by a factor $\hbar/|e|B$.

It follows from (\ref{ons}) that e.g. the
magneto-oscillations are periodic in $1/B$ with the periods determined by
$A_k$.  

Here we rewrite the Onsager-Lifshitz rule in a form which employs projections
$A_{k,z}$ of extremal cross-sections.  The reason is
that this form (of course equivalent to (\ref{ons})) is more
appropriate for comparison with the
quantum\discretionary{-}{-}{-}mechanical 
treatment described in the next section.

Let us denote by $\varphi$ the angle between the growth direction and
the direction of the magnetic field, then $\vec{B}\equiv (0, B
\sin\varphi, B \cos\varphi)$.  Multiplication of equation (\ref{ons})
by $\cos\varphi$ leads to the expression
\begin{equation}
A_{k,z} = \frac{2\pi|e|B_z}{\hbar}(n+\textstyle\frac{1}{2}),
\label{onsz}  
\end{equation}
in which the total field $B$ was replaced by the component $B_z$ and
the cross-section area $A_k$ of a Fermi surface by its projection to the
plane $k_z=0$, denoted by $A_{k,z}$. Similarly, multiplication of
equation (\ref{ons}) by $\sin\varphi$ leads to the relation between
the component $B_y$ and the projection $A_{k,y}$ of $A_k$ to the plane
$k_y=0$.

With the energy spectrum given by (\ref{disp}), the projection $A_{k,z}$  
of the cross-sections to the plane $z=0$ can be written as
\begin{equation}
A_{k,z}=\frac{\sqrt{2m}}{\hbar}2 \int 
\sqrt{E-\frac{\hbar^2k_y^2}{2m}+2t\cos(k_yd_y - k_{z0}d_z)} \,dk_y,
\label{akz}
\end{equation}
where $d_y=(B_y/B_z)d_z$.

\begin{figure}[tbh]
\includegraphics[width=\linewidth]{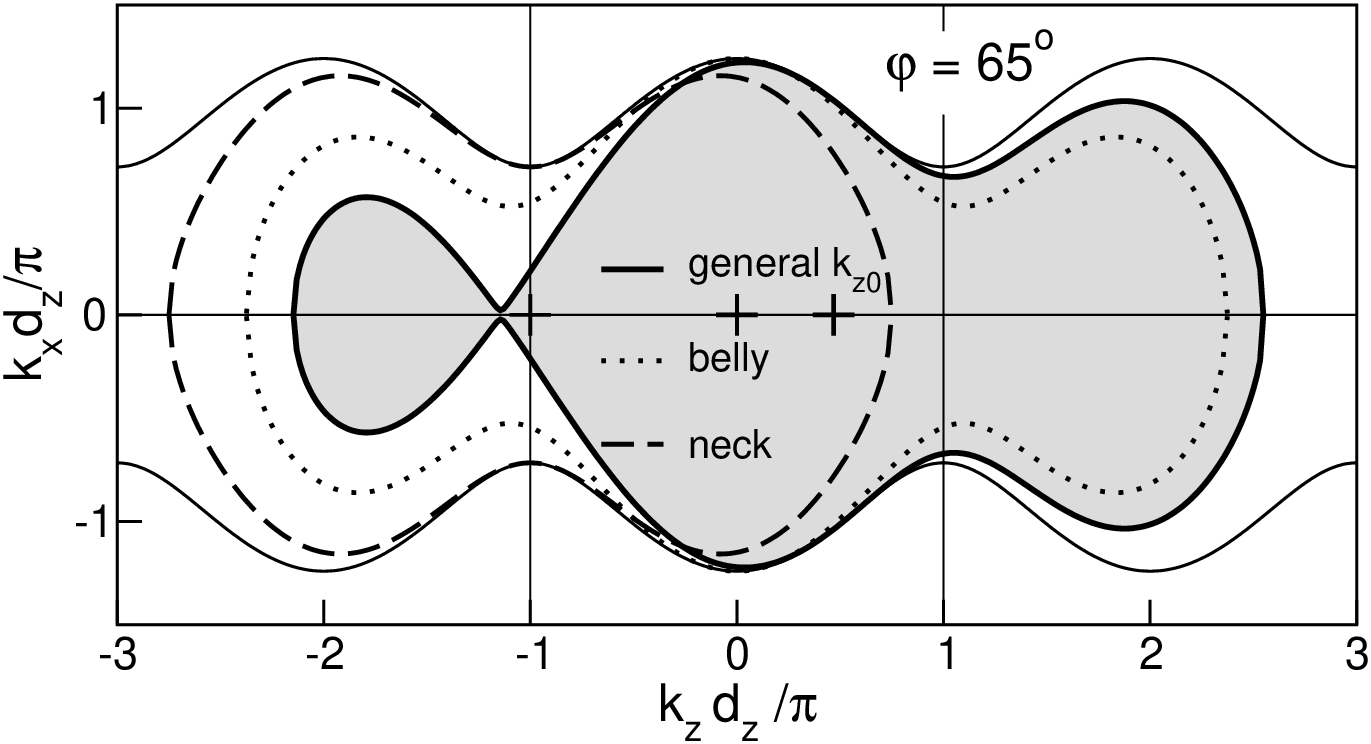}
\caption{\label{Fig2} Projections $A_{k,y}$ of three cross-sections of
the corrugated cylinder corresponding to $E_F=$ 15 meV (see
Fig.~\ref{Fig1}). Except of the ``belly'' and ``neck'' positions a
general $k_{z0}$ is considered, close to the critical value for which
the sudden drop of the area occurs. }
\end{figure}

Examples of projections $A_{k,y}$ corresponding to a ``belly'' and a
``neck'' cross-sections of the corrugated cylinder are shown in
Fig.~\ref{Fig2}. In that case two periods of magneto-oscillations exist.
The contribution of orbits corresponding to $k_{z0}$ between
$-\pi/d_z$ and $\pi/d_z$ to the oscillation amplitude is in general
weaker, except for special cases of ``extended'' orbits for
certain shapes of the Fermi surfaces and tilt angles. Note that a
sudden step of the cross-section area, shown in Fig.~\ref{Fig2} for a
general $k_{z0}$, can occur also for the extremal ``belly'' position,
if the field is slightly tilted from 65$^{\circ}$ towards the
perpendicular field configuration, as obvious from an inspection of
Fig.~\ref{Fig1}.

A single period corresponds to the Fermi surface formed by disconnected
ovals.  For independent electron layers (a smooth Fermi cylinder) the
equation (\ref{ons}) yields the energy spectrum $E_n= \hbar\omega_z(n+
\textstyle\frac{1}{2})$,where $\omega_z = |e|B_z/m$, for any tilt
angle. In that case the ``belly'' and ``neck'' areas are identical and
only one oscillation period exists.
%^^^^^^^^^^^^^^^^^^
%^^^^^^^^^^^^^^^^^^ 

%%%%%%%%%%%%%%%%%%%%%%%%%%%%%%%%%%%%%%%%%
\subsection{Quantum-mechanical approach\label{Quant}} 
%%%%%%%%%%%%%%%%%%%%%%%%%%%%%%%%%%%%%%%%%
The simple tight-binding model described in Sec. \ref{Tight} 
%on page~\pageref{Tight} 
can be generalized for the case of magnetic fields of arbitrary 
magnitude and orientation.

\begin{figure}[htb]
\includegraphics[width=\linewidth]{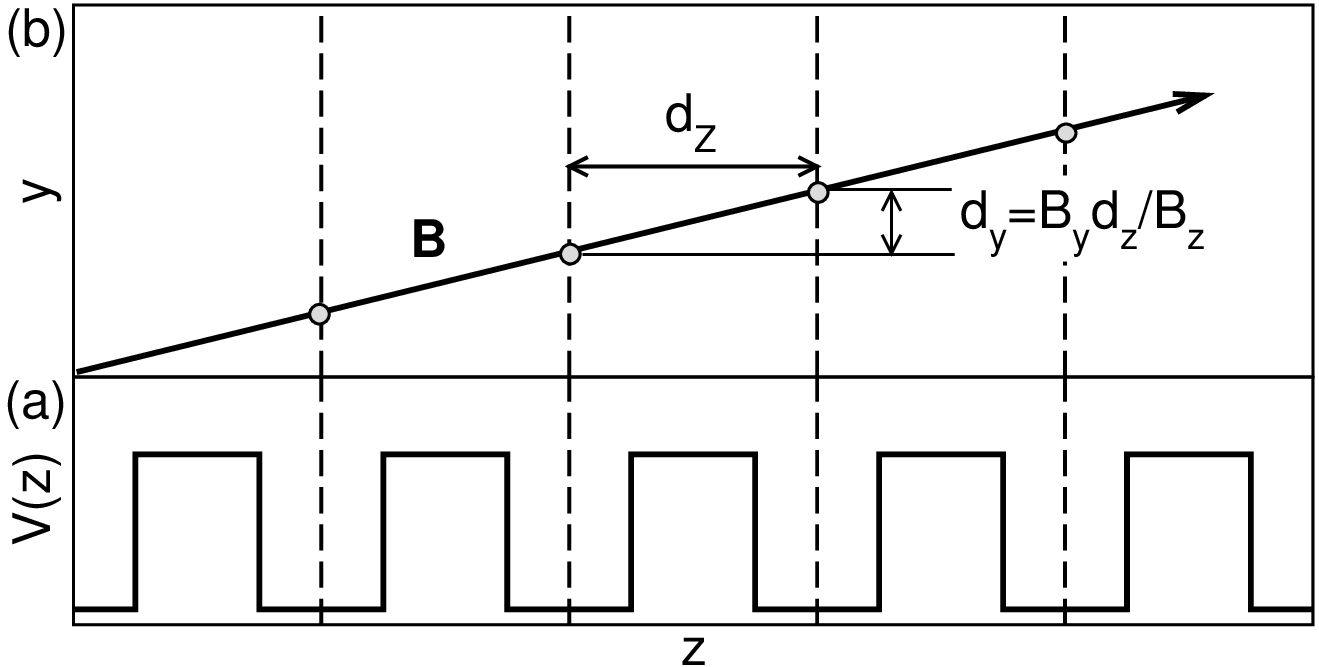}
\caption{\label{Fig3}(a) The schematic view of the superlattice
potential.  (b) The coordinates of minima of the ``electro-magnetic''
confining potential. An electron is bound by $V_b(z-Z_j)$ in 
the $z$-direction and by the magnetic harmonic potential $e^2(B_y Z_j
-B_zy)^2/2m$ in the $y$-direction. }
\end{figure}

Let us consider the superlattice subjected to a tilted magnetic field
$\vec{B}\equiv (0, B_y, B_z)$ given by the vector potential $\vec{A}
=(B_yz-B_zy,0,0)$. The 3D Hamiltonian $H$ of such a system,
\begin{equation}
H = \frac{1}{2m}\left(\vec{p}-e\vec{A}\right)^2 + V(z),
\end{equation}
depends on the variable $x$ only through the momentum component $p_x$
and, consequently, the corresponding
3D wave function
can be written in the form $\exp(ik_xx)\Phi(y,z)$. The function
$\Phi(y,z)$ is the solution to the
2D Schr\"odinger
equation with the Hamiltonian
\begin{equation}
H_{y,z} = \frac{p_y^2 + p_z^2}{2m} +
\frac{1}{2m} \left(\hbar k_x +|e|(B_yz - B_zy)\right)^2+ V(z).
\label{ham}
\end{equation}
This expression describes a linear array of quantum dots with the
minima of their potential energy at cross-sections of the lines 
\begin{eqnarray}
   \hbar k_x +|e|(B_yz - B_zy) & = & 0 , \\
                             z & = & Z_j,\nonumber
\end{eqnarray}
as shown in Fig.~\ref{Fig3}.  The coordinates of minima are given by
$\vec{R_j} = j\vec{d}$, where vector $\vec{d} \equiv (d_y, d_z)$ and
$d_y=B_y/B_zd_z$. The distance between two minima is
$d=\sqrt{d_y^2+d_z^2}$.
%
%^^^^^^^^^^^^^^^^^^
%^^^^^^^^^^^^^^^^^^

The eigenenergies of the corresponding Schr\"{o}dinger equation are
degenerated in $k_x$, the resulting $k_x$-degeneracy is
$|e|B_z/h$. The choice of $k_x$ means only an unessential shift of the
origin of coordinates $y$ or $z$, and we can set $k_x=0$ without lost
of generality.

We further assume, in agreement with Hu an MacDonald \cite{Hu}, that
the electron in an isolated well is still described by
$\chi_b(z-Z_j)$, as in the zero-magnetic-field case. The main effect
caused by the application of $\vec{B}$ is the restriction of the in-plane
motion of electrons in the $y$-direction by the parabolic ``magnetic''
potential with the center at $y=Y_j$. Thus, the zero-field plane wave
$\exp(ik_yy)$ should be replaced by a localized wave function which we
denote $\phi(y-Y_j)$.
%^^^^^^^^^^^^^^^^^^
\begin{figure}[b]
\includegraphics[width=\linewidth]{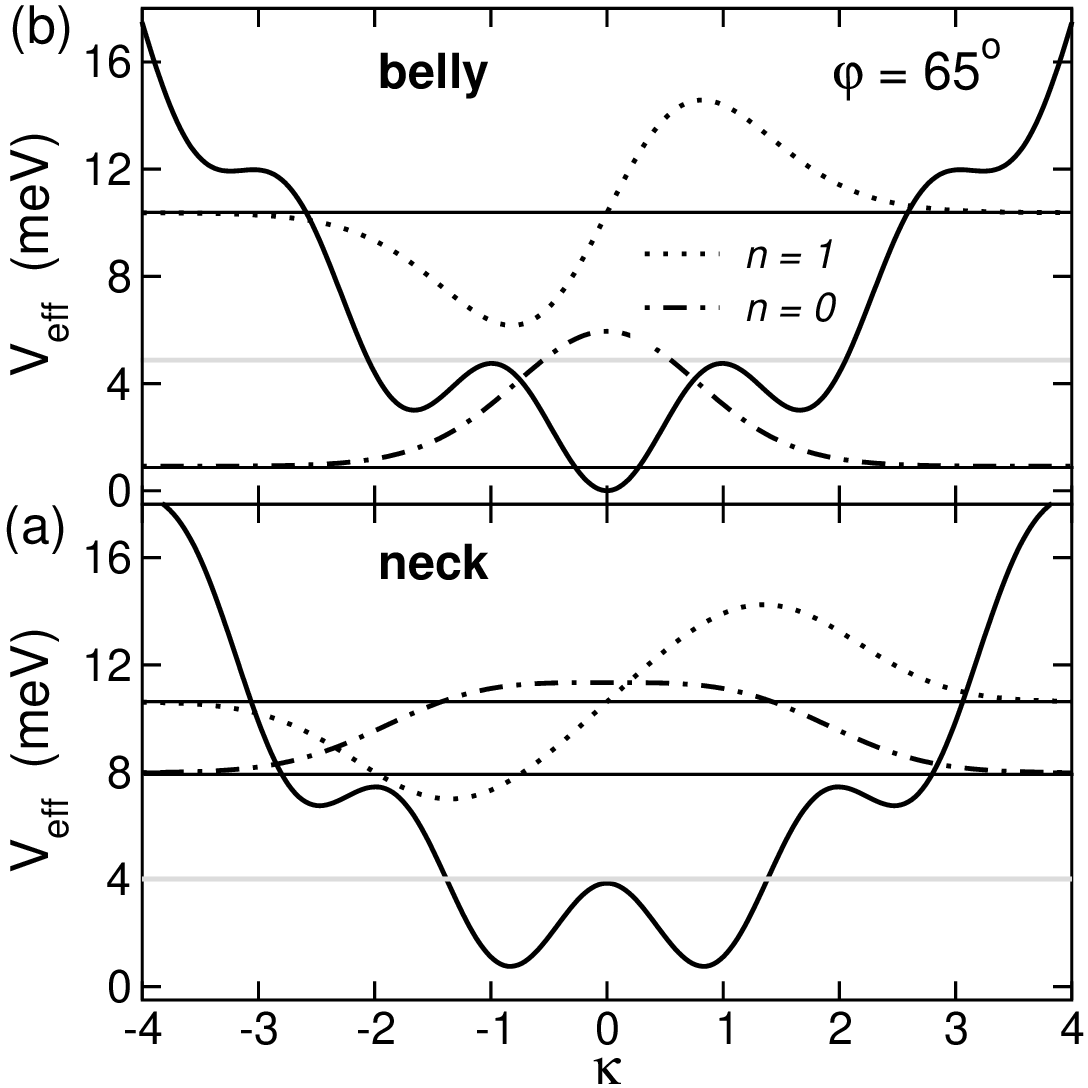}
\caption{\label{Fig4} The modulated parabolic well in $p$-space for
(a) the ``neck'' position and (b) the ``belly'' position of $k_{z0}$. Two
lowest eigenstates and eigenenergies are calculated for $B_z = 3$ T.
The grey lines  correspond to the critical parameters of
quasi\discretionary{-}{-}{-}classical orbits, 
at which the area calculated using (\ref{akz})
changes abruptly. The dimensionless variable $\kappa$ is determined
by $\hbar \kappa = \ell_z p_y$, where $\ell_z=\sqrt{\hbar/|e|B_z}$.} 
\end{figure}
%^^^^^^^^^^^^^^^^^^
Since the Hamiltonian (\ref{ham}) is periodic with the period $d$, we
can write the approximate wave function $\Phi(y,z)$ in the form of a
Bloch sum
\begin{equation}
\Phi(y,z) =
\sum_j e^{i\vec{k}\vec{R_j}} \chi_b(z-Z_j)\phi(y-Y_j),
\end{equation}
where $\vec{k}\equiv (k_y, k_z)$ is a wave vector oriented in the
$\vec{d}$-direction. The magnitude of the wave vector,
$k=\sqrt{k_y^2+k_z^2}$, varies between $-\pi/d$ and $\pi/d$, the
borders of the 1D Brillouin zone.

The matrix equation
\begin{equation}
\langle\chi_b(z-Z_j)|E-H_{y,z}|\Phi(y,z)\rangle =0
\end{equation}
yields the 1D Schr\"odinger 
equations from which the eigenfunctions $\phi_j=\phi(y-~Y_j)$ and the
corresponding eigenenergies are to be determined. Using
\begin{equation}
\phi(y+d_y) = \exp(i\frac{p_y}{\hbar}d_y)\phi(y),
\end{equation}
we obtain for each $j$
\begin{equation}
\left[\frac{p_y^2}{2m} +\frac{m\omega_z^2}{2}\left(y-Y_j\right)^2
-2t \cos\left(\frac{p_y}{\hbar}d_y-\vec{k}\vec{d}\right)\right]\phi_j
= E \phi_j.
\label{oned}
\end{equation}
As these equations are independent and equivalent for all values of
$j$, we can limit ourselves to a single equation with e.g. $j=0$.
Employing the $p$-representation, the equation (\ref{oned}) can be
written as
\begin{equation}
\left[-\frac{m\hbar^2\omega_z^2}{2}\frac{\partial^2}{\partial p_y^2}
+ \frac{p_y^2}{2m}
-2t \cos\left(\frac{p_y}{\hbar}d_y-\vec{k}\vec{d}\right)\right]\phi_0
= E \phi_0,
\label{onedp}
\end{equation}
where $\phi_0$ is a function of $p_y$, $\phi_0 = \phi_0(p_y)$.  Thus
the 3D Schr\"{o}dinger equation 
is reduced to 1D, with the 
energy spectrum formed by 1D Landau subbands 
$E_n(\vec{k})$.  This is the central result of this paper.

The parabolic well in $p$-space modulated by the cosine potential is
shown in Fig.~\ref{Fig4} for two values of the phase factor, $0$ and
$\pi$. The choice $\vec{k}\vec{d}=0$ corresponds to the
``belly'' orbit and $\vec{k}\vec{d}=\pi$ to the ``neck'' orbit in the
quasi\discretionary{-}{-}{-}classical terminology. 

In principle, the equation~(\ref{onedp}) can be solved
quasi\discretionary{-}{-}{-}classically (by the WKB method) 
or quantum\discretionary{-}{-}{-}mechanically. 
The choice of the method depends on the system parameters.

As anticipated, the quasi\discretionary{-}{-}{-}classical 
solution leads to the expression
(\ref{akz}). Note that the phase factor $\vec{k}\vec{d}$ can be
replaced by $k_{z0}d_z$.  (The projection $k_{z0}$ of $\vec{k}$ to the
$k_z$-axis satisfies $k_{z0}\in (-\pi/d_z,+\pi/d_z)$.)  Generally, the
WKB method is applicable if $\hbar\omega_z \ll 4t$ and many states
below the Fermi level are occupied.

The grey lines shown in Fig.~\ref{Fig4} denote tops of ``potential''
barriers which separate the classicaly inaccesible regions of
$p_y$.  The magnetic breakdown theory describes tunneling between two
{\em quasi\discretionary{-}{-}{-}classical} orbits from neighboring regions.

In semiconductor superlattices $\hbar\omega_z$ becomes comparable to
$4t$ in relatively weak magnetic fields. Two lowest eigenstates
calculated quantum\discretionary{-}{-}{-}mechanically are shown 
in Fig.~\ref{Fig4}. The wave functions extend over several 
local minima of the ``potential'', just in opposition to 
requirements of quasi\discretionary{-}{-}{-}classical 
approximation and the magnetic breakdown theory, which clearly 
cannot be applied in this case.

%%%%%%%%%%%%%%%%%%%%%%%%%%%%%%%%%%%%%%%%%%%%%%%%%%%%%%%%%%%%%%%%%%%%%%%%%%
\section{In-plane magnetic field \label{In-plane}}
%%%%%%%%%%%%%%%%%%%%%%%%%%%%%%%%%%%%%%%%%%%%%%%%%%%%%%%%%%%%%%%%%%%%
The above approach fails for the case $B_z\rightarrow 0$.
In an in-plane magnetic field $\vec{B}=(0,B_y,0)$  the vector 
potential takes the form $\vec{A} = (B_yz,0,0)$ and the
one-electron Hamiltonian (\ref{ham}) reduces to
\begin{equation}
H_{y} = \frac{p_y^2 + p_z^2}{2m} +
\frac{1}{2m} \left(\hbar k_x +|e|B_yz \right)^2+ V(z).
\label{hamy}
\end{equation}
Treating the inter-well hopping in a tight-binding approximation as in
Sec. \ref{Quant}, the Hamiltonian (\ref{hamy}) transforms to a 
three-diagonal matrix with diagonal and off-diagonal elements given by
\begin{equation}
\label{H_sl_diagonal}
H_{j,j}=\frac{\hbar^2}{2m}\left(k_x+k_j\right)^2+
\frac{\hbar^2k_y^2}{2m}, \qquad H_{j,j\pm 1}=-t,
\end{equation}
where $k_j =j K_0$ is the magnetic-field-dependent wave-vector with
$K_0=|e|B_yd_z/\hbar=d_z/\ell_y^2$.  This is a matrix form of the
Mathieu equation (see e.g. \cite{math}).

Solving the eigenvalue problem we get a number of Landau subbbands
$E_n(k_x)$ which are $K_0$--periodic in $k_x$. Then  the full energy
spectrum is given by
\begin{equation}
E_n(k_x,k_y) = E_n(k_x) +\frac{\hbar^2k_y^2}{2m}.
\label{spec}
\end{equation}
%
%^^^^^^^^^^^^^^^^^^
\begin{figure}[t]
\includegraphics[width=\linewidth]{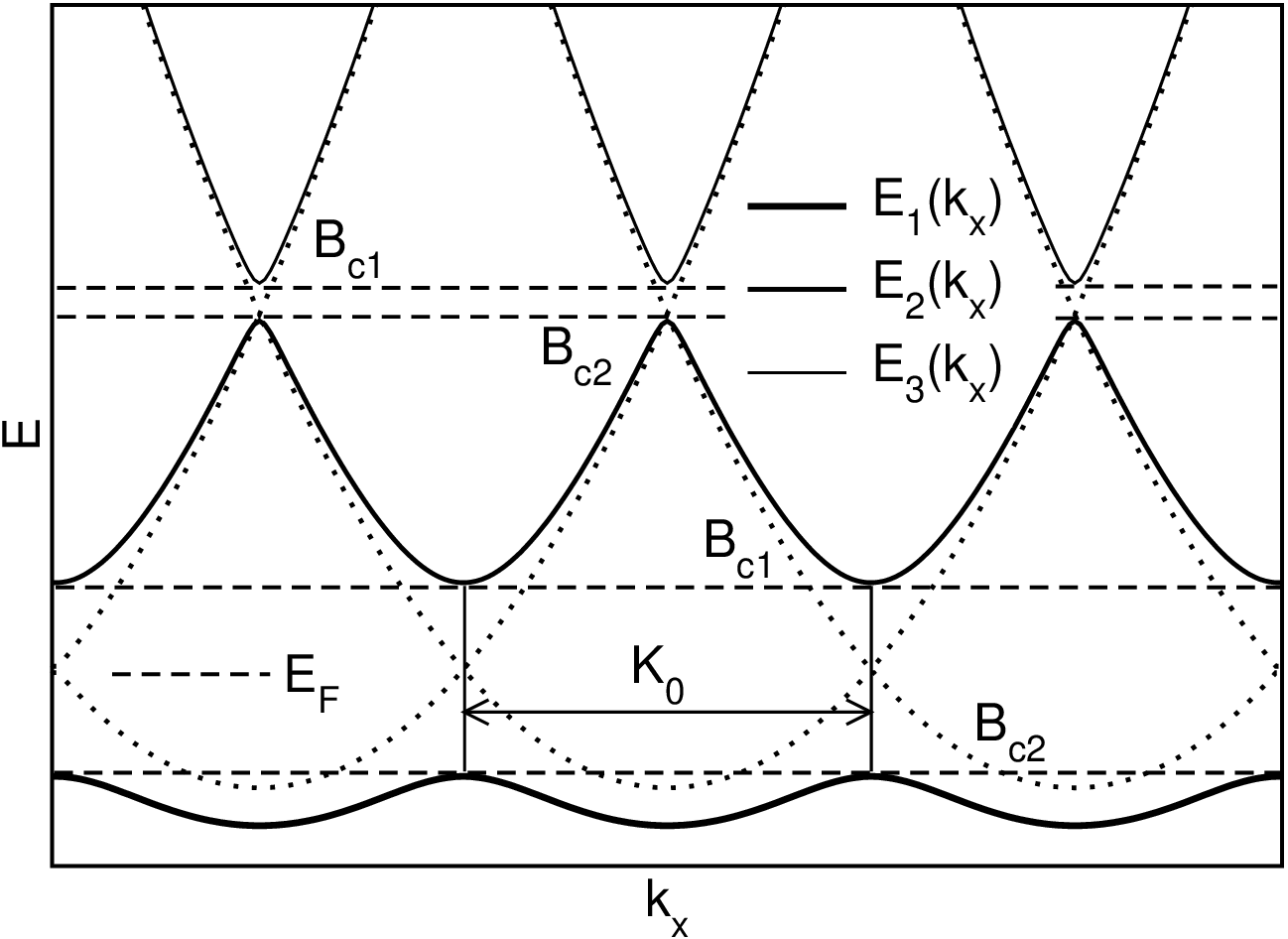}
\caption{\label{E_kx} The energy dispersion curves $E_n(k_x)$.
The dotted lines correspond to independent 2D electron layers, $t=0$. 
The dashed lines denote subband boundaries.} 
\end{figure}
%^^^^^^^^^^^^^^^^^^
The lowest subbands $E_n(k_x)$ are shown in Fig.~\ref{E_kx}.  The
states from neighboring layers, described by dotted parabolas, are
mixed at the border of Brillouin zones where the free-electron
parabolas cross.  The electrons tunnel between wells  near the
cross-points and energy gaps are opened there.

This is reflected in the shape of equi-energetic lines
$E_F=E_n(k_x,k_y)$.  For large enough $B_y$ the separation of centers,
$K_0$, becomes  larger than the diameter of the free-electro Fermi
circles $2k_F$. The contours do not cross and electrons cannot tunnel
at all.  Note that this condition is equivalent to $d_z > 2\ell_yk_F$,
proposed by Dingle \cite{Dingle} as mentioned in the introduction. 

With lowering the field (and smaller $K_0$) the Fermi contours first
``kiss'' on borders of Brillouin zones when $E_F$ touches the top of
the lowest Landau subband.  Then they merge into an open contour and a
new Fermi oval, belonging to the second subband, appears when $E_F$ reaches 
its bottom, etc.

This is illustrated by dashed lines in Fig.~\ref{E_kx}. For simplicity
we use fixed $B_y$ and variable $E_F$, instead of the fixed Fermi
energy and sweeping $B_y$. 

Two types of critical magnetic fields (energies) can be distinguished,
$B_{c1}$ and $B_{c2}$.  When the Fermi energy touches the bottom of a
Landau subband at $B_{c1}$ , there is the minimum in $E_n(k_x,k_y)$
and the step van Hove singularity appears in the density of states. The
second possibility is the Fermi energy coinciding with the maximum of
an $E_n(k_x)$ at $B_{c2}$. It corresponds to the saddle point in
$E_n(k_x,k_y)$ and the logarithmic van Hove singularity in the density
of states.

%%%%%%%%%%%%%%%%%%%%%%%%%%%%%%%%%%%%%%%%%%%%%%%%%%%%%%%%%%%%%%%%%%%%%%%%%%
\section{Numerical example \label{Num}}
%%%%%%%%%%%%%%%%%%%%%%%%%%%%%%%%%%%%%%%%%%%%%%%%%%%%%%%%%%%%%%%%%%%%
It is clearly illustrated in Fig.~\ref{Fig4} that the
quantum\discretionary{-}{-}{-}mechanical regime, when only 
a few Landau subbands are occupied
and the quasi\discretionary{-}{-}{-}classical approach 
is not valid, can be reached in
relatively week magnetic fields for realistic parameters of
superlattices.  In the quasi\discretionary{-}{-}{-}classical 
approach the electron motion in the effective ``potential'' 
should be limited by local side-maxima,
due to the cosine modulation. In quantum mechanics electrons can tunnel
through the local barriers even if their energy is below the
maxima. In principle, it is possible to describe tunneling through
the barriers quasi\discretionary{-}{-}{-}classically as the 
``magnetic breakdown'', but it seems not very appropriate 
for states with lowest quantum numbers.

In experiments, usually the concentration of carriers is kept fixed
and the magnetic field is varied. In most cases experimental data
reflect the field-induced singularities of the density of
states. Therefore, we concentrate in our numerical example on the evaluation of
the density-of-states field dependence for a series of tilt
angles. The results of both approaches will be compared. 

For simplicity, we start with the analytically solvable case of
perpendicular magnetic fields.

\subsection{Perpendicular magnetic field} 
%%%%%%%%%%%%%%%%%%%%%%%%%%%%%%%%%%%%%%%%%
In the quasi\discretionary{-}{-}{-}classical approximation, 
two extremal circular orbits, a
``belly'' and a ``neck'', can be found on the corrugated cyllinder
with the cross-sections
\begin{equation}
A_{k, belly} = \pi\frac{2m}{\hbar^2}(E_F+2t), \,\,\,
A_{k, neck} = \pi\frac{2m}{\hbar^2}(E_F-2t).
\label{bellyneck}
\end{equation}  
For these orbits, the quantization condition~(\ref{ons}) can be rewritten 
as
\begin{equation}
E_F\pm2t = \hbar\omega_z(n+\textstyle\frac{1}{2})
\label{bn}
\end{equation}  
which yields two periods of oscillations
\begin{equation}
\Delta\left(\frac{1}{B_z}\right)_{\stackrel{belly}{neck}} =
\frac{\hbar|e|}{m(E_F\pm 2t)} .
\label{perbellyneck}
\end{equation}  
It follows from (\ref{perbellyneck}) that for the Fermi energy close
to the  miniband top the period of a ``neck'' orbit increases and
reaches infinity at  $E_F = 2t$. For $E_F$ within the miniband 
the Fermi surface consists of disconnected ovals  and only one
oscillation period exists, corresponding to the ``belly'' orbits.

The energy spectrum obtained by full
quantum\discretionary{-}{-}{-}mechanical solution of
the Schr\"odinger equation  profits from the fact that
in perpendicular magnetic fields $d_y=0$. Consequently, the equation 
(\ref{onedp})
is reduced to the standard equation of free electron motion in
perpendicular magnetic fields. Landau subbands are formed by
a sum of Landau energy levels and the
1D miniband, like in the zero-field case:
\begin{equation}
E_n(\vec{k}) = \hbar\omega_z(n+\textstyle\frac{1}{2})-2t\cos(k_zd_z).
\end{equation}  
Due to the periodic potential, the Landau levels are broadened
into Landau subbands.

The corresponding density of states per layer, $g(E)$,   can be
evaluated analytically and reads
\begin{equation}
g(E) = \frac{|e|B_z}{h}\frac{1}{\pi} \sum_n \frac{1}
{\sqrt{4t^2-(E-\hbar\omega_z(n+\textstyle \frac{1}{2}))^2}}.
\end{equation}  
Two van Hove singularities (of the type $1/\sqrt E$) are due to the
maximum and minimum of the Landau subband at the borders of the
Brillouin zone, their positions on the energy axis are given by the
equation (\ref{bn}), i.e. they correspond to the extremal ``belly''
and ``neck'' orbits obtained quasi\discretionary{-}{-}{-}classically. 
At fixed Fermi energy, these extrema define two oscillation periods in
the $B_z$-dependency of the density of states.
%They determine two periods of oscillations if we keep the 
%Fermi energy fixed and consider the density of states 
%as a function of $B_z$.
%
%^^^^^^^^^^^^^^^^^^
\begin{figure}[b]
\includegraphics[width=\linewidth]{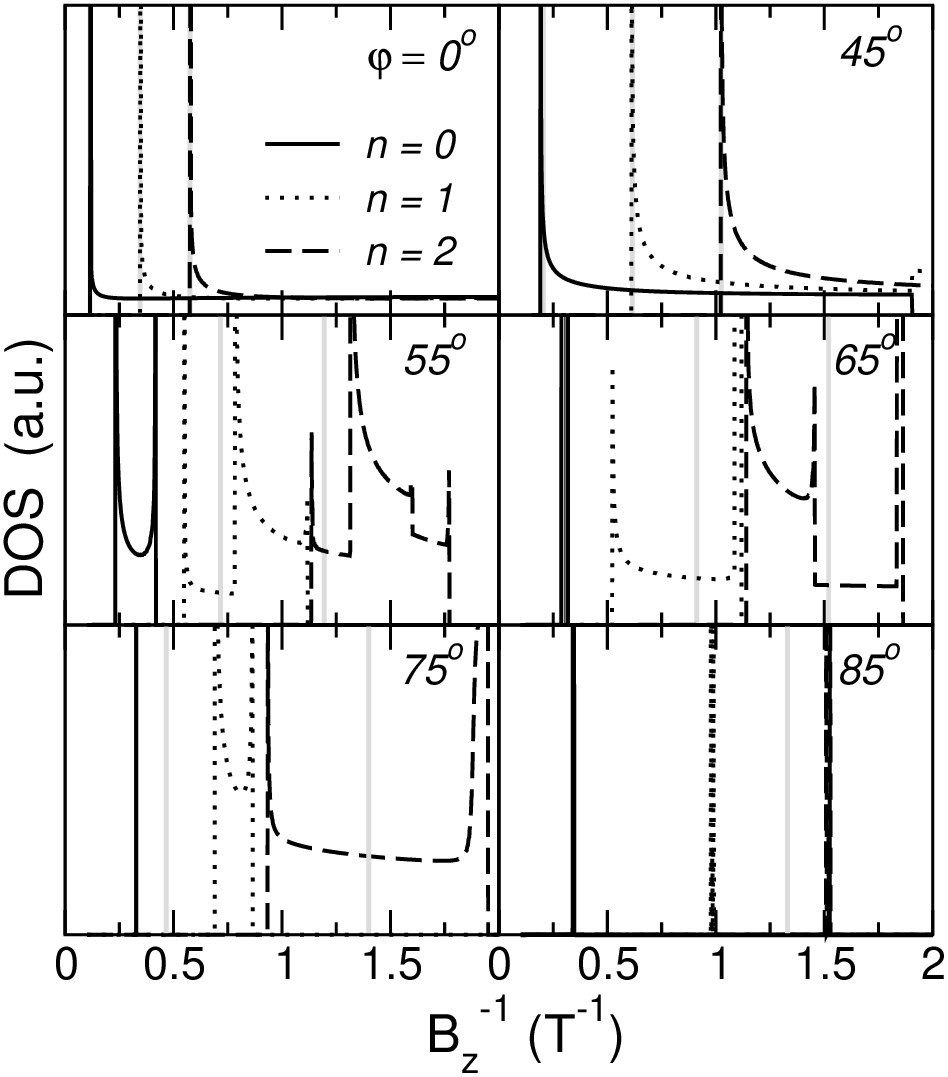}
\caption{\label{Fig5} 
The magnetic-field-dependent density of states calculated for Fermi
energy 7.5 meV above the miniband bottom. First three Landau subbands are
shown. The grey lines correspond to singularities calculated
quasi\discretionary{-}{-}{-}classically. }
\end{figure}
%^^^^^^^^^^^^^^^^^^

%%%%%%%%%%%%%%%%%%%%%%%%%%%%%%%%%%%%%%%%%%%%%%%%%%%%%%%%%%
\subsection{Tilted magnetic fields}
%%%%%%%%%%%%%%%%%%%%%%%%%%%%%%%%%%%%%%%%%%%%%%%%%%%%%%%%%
To stress the difference between the quasi\discretionary{-}{-}{-}classical and 
quantum\discretionary{-}{-}{-}mechanical solutions, 
we shall consider a superlattice with the 
closed semi-elliptic Fermi surface, presented in Fig.~\ref{Fig1}. 

In such a case the quasi\discretionary{-}{-}{-}classical solution 
leads to single-period oscillations, with the period 
depending on the area of the ``belly''
extremal cross-sections, which vary with the tilt angle. 
With the distance of non-extremal to the extremal orbits their
contribution to the oscillation amplitude smoothly vanishes.
%The contribution of non-extremal orbits to the oscillation amplitude
%smoothly vanishes with their distance from the extremal orbit.

As the largest difference between the quasi\discretionary{-}{-}{-}classical and
quantum\discretionary{-}{-}{-}mechanical approach is 
expected for lowest eigenenergies, we
present in Fig.~\ref{Fig5} the density of states of three lowest
Landau subbands calculated for the fixed $E_F$ as a function of
$1/B_z$. To study the increasing influence of the in-plane field
component $B_y$, the tilt angle is varied from the perpendicular
position, $\varphi =0$, towards the in-plane field configuration. In
addition to the curves obtained from the
quantum\discretionary{-}{-}{-}mechanical 
solution of equation (\ref{onedp}), the position of 
the quasi\discretionary{-}{-}{-}classical ``belly'' orbits is shown.

As mentioned above, both methods yield exactly the same results in the
perpendicular magnetic field orientation, $\varphi = 0$, and
only small deviations of singularities corresponding to the
``belly'' orbits are found for angles up to $45^{\circ}$.  At
$45^{\circ}$ marked deviations appear, the ``neck'' singularity
returns to the first Landau subband, i.e. its width becomes
finite. Moreover, an additional singularity occurs in the second
Landau subband.
%
%^^^^^^^^^^^^^^^^^^
\begin{figure}[htb]
\includegraphics[width=\linewidth]{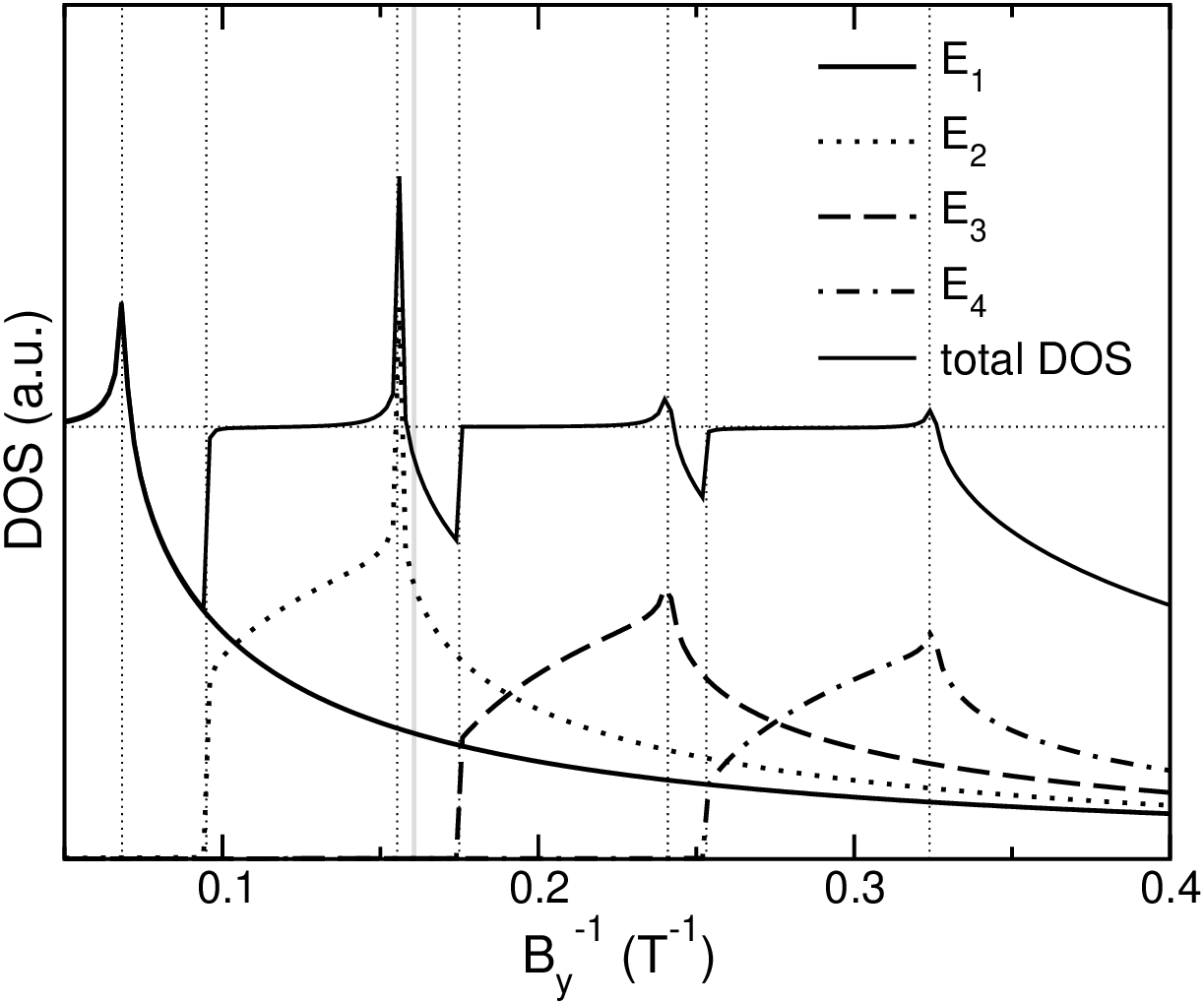}
\caption{\label{Fig7} 
The magnetic-field-dependent density of states calculated for Fermi
energy 7.5 meV above the miniband bottom. The grey line corresponds to
the singularity calculated quasi\discretionary{-}{-}{-}classically. 
(Only one singularity is depicted.) }
\end{figure}
%^^^^^^^^^^^^^^^^^^ 
Above $45^{\circ}$ the two solutions are completely different. While
the period of quasi\discretionary{-}{-}{-}classical 
oscillations monotonously increases with
the decreasing area of the ``belly'' cross-section, the quantum
mechanical solution exhibits new features. Two singularities in the
first Landau subband become closer as $\varphi$ grows, the width
of the subband shrinks and around $65^{\circ}$ it reminds a level.

In the second Landau subband,
besides the ``belly'' and ``neck'' singularities an additional
singularity shows up between them which
vanishes above $65^{\circ}$. Then also the second subband shrinks in 
a level.

Four singularities appear in the third subband, for higher $\varphi$
their number is reduced to three and two before the subband shrinks to
a level. Obviously, the number of singularities is related to the
number of nodes of the corresponding wave functions.

%%%%%%%%%%%%%%%%%%%%%%%%%%%%%%%%%%%%%%%%%%%%%%%%%%%%%%%%%%%%%%%
\subsection{In-plane magnetic fields\label{}}
%%%%%%%%%%%%%%%%%%%%%%%%%%%%%%%%%%%%%%%%%%%%%
For the semi-elliptic Fermi surface there is no difference between the
in-plane and tilted magnetic field in the quasi\discretionary{-}{-}{-}classical
approximation.  The single period of oscillations is determined by the
extremal cross-section which is now in the $k_x k_z$-plane. The
correspoding singularity is shown in Fig.~\ref{Fig7} together with the
density of states which results from processing the numericaly
obtained eigenenergies $E_n(k_xk_y)$. Both results show a pronounced
qualitative difference. 
  
The quasi-clasical approach does not distinguish between the
perpendicular and in-plane magnetic fields.  It assumes that the
1D subbands attached to the 
quasi\discretionary{-}{-}{-}classicaly calculated
Landau levels are emptied by increasing the magnetic field. But this is
correct only for the perpendicular field orientation.

In the quantum\discretionary{-}{-}{-}mechanical picture 
2D Landau subbands 
are emptied and instead of the van Hove singularities of 
$1/\sqrt(E)$ type logarithmic and step singularities appear.

%%%%%%%%%%%%%%%%%%%%%%%%%%%%%%%%%%%%%%%%%%%%%%%%%%%%%%%%%%%%%%%
\section{Conclusions\label{Concl}}
%%%%%%%%%%%%%%%%%%%%%%%%%%%%%%%%%%%%%%%%%%%%%%%%%%%%%%%%%%
In superlattices with the period $d_z$, the in-plane magnetic field
component $B_y$ displaces the origins of the Fermi circles of
neighboring electron layers by $|e|B_y d_z/\hbar$. Above the critical
field $B_{y,c}=2\hbar k_F/|e|d_z$ the Fermi circles with the radii
$k_F$ cannot cross and tunneling is impossible.

The theoretical description is particularly simple for the
tight-binding model of the electronic structure; the generalized
Landau eigenenergies in tilted magnetic fields can be found as
solutions to a one\discretionary{-}{-}{-}dimensional Schr\"{o}dinger equation.

The quasi\discretionary{-}{-}{-}classical solution 
to this 1D problem yields the standard 
Onsager-Lifshitz rule. The full
quantum\discretionary{-}{-}{-}mechanical solution is necessary to describe the
3D$\rightarrow$2D transition, i.e. the transition to a sequence of
independent 2D electron layers.
%%%%%%%%%%%%%%%%%%%%%%%%%%%%%%%%%%%%%%%%%%%%%%%%%%%%%%%%%%%%%%%%%%%%%%%%%
\section{Acknowledgements}
%%%%%%%%%%%%%%%%%%%%%%%%%%%
This work has been supported by the Grant Agency of the ASCR under
Grant No. IAA1010408, by the French-Czech Project Barrande
2003-013-2 and by the Institutional Research Plan No. AV0Z10100521.


\begin{thebibliography}{9}

\bibitem{Esaki} 
L.\ Esaki and R.\ Tsu, IBM J.\ Res.\ Dev. {\bf 14}, 61 (1970).

\bibitem{maan}
J.\ K.\ Maan, Adv.\ Solid State Phys.\ {\bf 27}, 137 (1987).

\bibitem{Moses1} 
R.\ H.\ McKenzie and P.\ Moses, Phys.\ Rev.\ Lett.\ {\bf 81},  4492 (1998).

\bibitem{Moses2} 
P.\ Moses and R.\ H.\ McKenzie, Phys.\ Rev.\ B {\bf 60},  7998 (1999).

\bibitem{Yamaji} K.\ Yamaji, J.\ Phys.\ Soc.\ Jpn.\ {\bf 58},  1520 (1989).

\bibitem{Osada1} 
T.\ Osada, H.\ Nose, and M.\ Kuraguchi, Physica B  {\bf 294--295},
 402 (2001).

\bibitem{Osada2}
T.\ Osada, Physica E {\bf 12},  272 (2002).

\bibitem{Osada3}
T.\ Osada, M.\ Kuraguchi, K.\ Kobayashi, and  E.\ Ohmichi,
Physica E {\bf 18}, 200  (2003).

\bibitem{Stormer}
H.\ L.\ St{\"o}rmer, J.\ P.\ Eisenstein, A.\ C.\ Gossard, 
W.\ Wiegmann, and K.\ Baldwin, 
Phys. Rev. lett. {\bf 56},  85 (1986).

\bibitem{on} L.\ Onsager, Phil.\ Mag.\ {\bf 43}, 1006 (1952).

\bibitem{Ashcroft} 
N.\ W.\ Ashcroft and N.\ D.\ Mermin, in {\it Solid State Physics} 
(Saunders, Philadelphia, 1975), p.\ 368.

\bibitem{Eaves1}
C.\ R.\ Tench, T.\ M.\ Fromhold, S.\ Bujkewicz, P.\ B.\ Wilkinson,
F.\ W.\ Sheard, and  L.\ Eaves, 
Physica B {\bf 272},  209 (1999).

\bibitem{Eaves2}
S.\ Bujkewicz, T.\ M.\ Fromhold, M.\ J.\ Carter, F.\ W.\ Sheard, and L.\ Eaves,
Physica E {\bf 7},  827 (2000).

\bibitem{Osada4} 
K.\ Kobayashi, M.\ Saito, E.\ Ohmichi, and T.\ Osada, 
Physica E {\bf 22}, 385 (2004).

\bibitem{Eaves3}
 T.\ M.\ Fromhold, A.\ Patan\`{e},  S.\ Bujkewicz,  P.\ B.\ Wilkinson,
D.\ Fowler, D.\ Sherwood, S.\ P.\ Stapleton, A.\ A.\ Krokhin,  L.\ Eaves, 
M.\ Henini, N.\ S.\ Sankeshwar, and  F.\ W.\ Sheard,
Nature {\bf 428} 726 (2004).
 
\bibitem{Falicov}
M.\ H.\ Cohen and L.\ M.\ Falicov, Phys.\ Rev.\ Lett.\ {\bf 7}, 231 (1961).

\bibitem{Blount}
E.\ I.\ Blount, Phys.\ Rev.\ {\bf 126}, 1636 (1961).

\bibitem{Chambers}
W.\ G.\ Chambers, Proc.\ Phys.\ Soc.\ {\bf 82}, 181 (1964).

\bibitem{Jaschinski} 
O.\ Jaschinski, G.\ Nachtwei, J.\ Schoenes, P.\ B\"onsch, 
and A.\ Schlachetzki, Physica  B {\bf251}, 873 (1998).

\bibitem{Nachtwei}
G.\ Nachtwei, A.\ Weber, O.\ Jaschinski, and H.\ K\"unzel, 
J.\ \nolinebreak Appl.\ Phys.\ {\bf84}, 323 (1998).

\bibitem{Kawamura_1}
%"Quantum Hall Effect in GaAs/AlGaAs Semiconductor Superlattice",
M.\ Kawamura, Doctoral Thesis (University of Tokyo) December 2000, p.\ 31.

\bibitem{Kawamura_2}
M.\ Kawamura, A.\ Endo, S.\ Katsumoto, Y.\ Iye, C.\ Terakura, S.\ Uji,
Physica B {\bf298}, 48 (2001).

\bibitem{Dingle} 
R.\ Dingle, Surf.\ Sci.\ {\bf 73}, 229 (1978).

\bibitem{Hu} 
J.\ Hu and A.\ H.\ MacDonald, Phys.\ Rev.\ B {\bf 46},  12554 (1992).

\bibitem{bastard} 
G.\ Bastard, in {\em Wave Mechanics Applied to Semiconductor
Heterostructures} (Monographies de physique, Paris, 1992), p.\ 22.  

\bibitem{math}
S.\ G.\ Davidson and M.\ St\'{e}\c{s}licka, 
in {\em Basic Theory of Surface States} (Claredon Press,Oxford, 1992),
p.\ 41.

\end{thebibliography}
\end{document}